\documentclass{aastex63}

\usepackage{amsmath}

\newcommand{\elsasser}{Els\"asser\ }
\newcommand{\alfven}{Alfv\'en\ }

\received{}
\revised{}
\accepted{}
\submitjournal{ApJ}

\shorttitle{Kink waves with Els\"asser variables}
\shortauthors{Van Doorsselaere et al.}

\begin{document}

\title{Wave pressure and energy cascade rate of kink waves computed with \elsasser variables}

\correspondingauthor{Tom Van Doorsselaere}
\email{tom.vandoorsselaere@kuleuven.be}

\author[0000-0001-9628-4113]{Tom Van Doorsselaere}
\affiliation{Centre for mathematical Plasma Astrophysics, Department of Mathematics, KU~Leuven, Celestijnenlaan 200B, B-3001 Leuven, Belgium}

\author{Bo Li}
\affiliation{Institute of Space Sciences, Shandong University, Weihai 264209, People's Republic of China}

\author{Marcel Goossens}
\affiliation{Centre for mathematical Plasma Astrophysics, Department of Mathematics, KU~Leuven, Celestijnenlaan 200B, B-3001 Leuven, Belgium}

\author{Bogdan Hnat}
\affiliation{Centre for Fusion, Space and Astrophysics, Department of Physics, University of Warwick, CV4 7AL, Coventry, UK}

\author{Norbert Magyar}
\affiliation{Centre for Fusion, Space and Astrophysics, Department of Physics, University of Warwick, CV4 7AL, Coventry, UK}



\begin{abstract}
	
Numerical simulations have revealed a new type of turbulence of unidirectional waves in a plasma that is perpendicularly structured \citep{magyar2017}, named {\em uniturbulence}. For this new type of turbulence, the transverse structuring modifies the upward propagating wave to have both \elsasser variables, leading to the well-known perpendicular cascade. In this paper, we study an analytical description of the non-linear evolution of kink waves in a cylindrical flux tube, which are prone to uniturbulence. We show that they lead to a non-linear cascade for both propagating and standing waves. We calculate explicit expressions for the wave pressure and energy cascade rate. 
The computed damping rate $\tau/P$ depends on the density contrast of the flux tube and the background plasma and is inversely proportional to the amplitude of the kink wave. The dependence on the density contrast shows that it plays a role especially in the lower solar corona. 
These expressions may be added in \alfven wave driven models of the solar atmosphere \citep[such as AWSOM,][]{vanderholst2014}, modifying it to UAWSOM (Uniturbulence and \alfven Wave Solar Model).

\end{abstract}



\section{Introduction} \label{sec:intro}
It is a long-standing belief that MHD turbulence develops only when two counterpropagating \alfven waves interact. The subsequent interaction leads to a perpendicular cascade \citep{goldreich1995,matthaeus1999}. The cascade is scale invariant \citep{kolmogorov1962} and leads to the formation of a power law. For an overview of this theory and the associated observational and modelling results, please see the review by \citet{bruno2005}. This theory has been extended to also incorporate weak turbulence \citep{chandran2009} or magnetic field expansion \citep{dmitruk2002}.\par

It has become very obvious in recent years that the corona is filled with transverse waves \citep{tomczyk2007,depontieu2007,mcintosh2011,anfinogentov2015}. This has been used as motivation to extend previous models of an \alfven wave driven solar corona \citep{moriyasu2004,buchlin2007,antolin2008} and solar wind \citep{suzuki2005}. Recent models use 2D descriptions with reduced MHD to model the propagation and non-linear interaction of \alfven waves \citep{vanballegooijen2011,verdini2019}. Moreover, the models with \alfven wave turbulence have been extended to 3D using a homogeneous background \citep{rappazzo2008} and compressible turbulence \citep{shoda2019}. However, as also pointed out in \citet{chandran2019}, none of these models (allow to) take into account the significant density variation across the magnetic field, and this probably plays a crucial role in the wave dynamics \citep{vd2008} including the introduction of phase mixing.\par

In the latest generation models for the solar atmosphere, \alfven waves are driven from the photosphere and their evolution is studied in a WKB fashion on 1D field lines in 3D models \citep{vanderholst2014}, and are thus called \alfven Wave Solar Model (AWSOM). They take into account the \alfven wave pressure in addition to the plasma pressure and the extra energy dissipation by the \alfven wave turbulence to drive and heat the solar wind. Because of its success, this AWSOM approach was also taken up in other models \citep{mikic2018}, and its effect is now considered for the propagation of CMEs in the heliosphere \citep{verbeke2019}. However, as pointed out by \citet{verdini2019} and van der Holst (private communication), these models lack wave heating in the polar regions with open magnetic field and it seems that a crucial ingredient is missing in open magnetic fields in order to drive the solar wind. \par

Recently, it was found that kink waves, with a highly Alfv\'enic nature \citep{goossens2009}, that propagate in one direction on a field-aligned flux tube, become turbulent, and this phenomenon was called {\em uniturbulence} \citep{magyar2017}. It is related to generalised phase mixing, in the sense that wave property variation plays a crucial role. The work of \citet{magyar2019} demonstrated that the uniturbulent behaviour is due to inhomogeneities modifying wave properties which then deviate from the local \alfven waves. So, while classical phase mixing \citep{heyvaerts1983} studies the effects of varying phase speed on wave eigenfunctions, uniturbulence takes into account the modified eigenfunction to study the wave's non-linear evolution. \citet{magyar2019} showed that the essential ingredients for uniturbulence are (1) field-aligned density structures, (2) wave perturbations with a polarisation with a component along the gradient of density, (3) a wave vector with a component perpendicular to both the magnetic field and density gradient. As such, they found that surface \alfven waves are the simplest waves that exhibit the uniturbulence phenomenon. In this case, the surface \alfven waves are shown to have both \elsasser components simultaneously and co-propagating. This allows for classic \alfven turbulence to take place leading to the self-cascade that is uniturbulence. It was shown by \citet{shoda2018b} that this generalised phase mixing is at least important in the lower solar atmosphere, where sufficient density contrast exists, and where not a lot of reflected \alfven wave power occurs. Forward models of uniturbulence \citep{pant2019} can at least explain self-consistently the observed correlation between spectral line width and Doppler shift \citep{mcintosh2012}.\par

In this paper, we aim to present an analytical formalism that describes the kink wave pressure and energy dissipation rate for field-aligned flux tubes with a given density contrast. These expressions can be taken along in models like AWSOM to serve as an extra heating and wind driving mechanism in the open field regions. These generalised AWSOM models can be called UAWSOM (Uniturbulence and \alfven Wave Solar Model). We focus on the description of the energy input at large scales and low frequencies, and thus we rather model the energy input into the turbulent processes at these scales than the turbulence itself. Our results may be related to the $k^{-1}$ power law that is found to gradually get more important the closer to the Sun we get.\par

The formalism can also be important for the description of the non-linear evolution of kink waves in coronal loops. It has been shown numerically that kink waves excite the KHI \citep{terradas2008b}, leading to the formation of Transverse Wave Induced Kelvin-Helmholtz rolls (TWIKH rolls). The KHI has potential for explaining amplitude dependent damping of kink waves \citep{goddard2016, nechaeva2019, magyar2016b}, formation of strand-like structures \citep{antolin2014}, and broad DEMs \citep{vd2018}. These models show good promise to explain the observations of decayless kink waves \citep{wang2012,nistico2013} in driven KHI models \citep{karampelas2018,karampelas2019}.

\section{Terminology \& Aims}\label{sec:elsasser}
We define the \elsasser variables $\vec{Z}^\pm$ in the usual way:
\[\vec{Z}^\pm=\vec{v} \pm \frac{\vec{B}}{\sqrt{\mu\rho}},\]
where $\vec{v}$ and $\vec{B}$ are the velocity and magnetic field of the plasma, $\rho$ is the density and $\mu$ is the magnetic permeability. The equations of incompressible MHD can be rewritten using the \elsasser variables. The resulting equation is \citep{marsch1987, vanderholst2014, magyar2019}:
\begin{equation}\frac{\partial \vec{Z}^\pm}{\partial t}+\vec{Z}^\mp \cdot \nabla \vec{Z}^\pm=-\frac{1}{8}\nabla (\vec{Z}^+-\vec{Z}^-)^2, \label{eq:fullelsasser}\end{equation}
which has not been linearised yet! \par
Let us now proceed to a splitting of the quantities in background quantities and perturbations \citep[or ``turbulent components'',][]{vanderholst2014} on top of a static equilibrium, most importantly $\vec{B}=\vec{B}_0+\vec{b}, \rho=\rho_0+\rho'$. The \elsasser variables are split up as
\begin{equation}
	\vec{Z}^\pm=\pm V_\mathrm{A}\vec{e}_z + \vec{z}^{\,\pm},
\end{equation}
where
\begin{equation}
	\vec{z}^{\,\pm}=\vec{v}\pm \left( \frac{\vec{b}}{\sqrt{\mu\rho_0}} - \vec{e}_z V_\mathrm{A} \frac{\rho'}{2\rho_0}\right),\label{eq:linearelsasser}
\end{equation}
where $V_\mathrm{A}=\frac{B_0}{\sqrt{\mu\rho_0}}$ is the usual \alfven speed. The last term in the expression is often missing in turbulence descriptions, because \alfven waves in a uniform medium as considered in those studies do not have associated density fluctuations. For kink waves, it is an essential term, because they are not entirely incompressible and do have associated density fluctuations. These compressible effects become larger with increasing wave number. \\ We can subtract the equilibrium component from Eq.~\ref{eq:fullelsasser} and obtain the following equation for the evolution of the turbulent part of the \elsasser variables:
\begin{equation}
	\frac{\partial \vec{z}^{\,\pm}}{\partial t}\mp V_\mathrm{A}\frac{\partial \vec{z}^{\,\pm}}{\partial z}=-\vec{z}^{\,\mp} \cdot \nabla \vec{z}^{\,\pm},\label{eq:linearmhd}
\end{equation}
where we have moved the non-linear terms to the right-hand side. \par
\citet{vanderholst2014} consider in their AWSOM model for the heliosphere the additional pressure $P_\mathrm{A}$ exerted by the \alfven waves, which are modelled in a 1D, WKB approximation along the field line. They also take into account the energy dissipation rate $\epsilon^\pm$ from the \alfven turbulence that is being cascaded to the smallest scales where it is dissipated and absorbed as thermal energy. Their expressions for these key quantities are (using the manipulations in their Eq.~20, 22 and 25, without assuming incompressibility):
\begin{equation}
	P_\mathrm{A}=\frac{\vec{b}^2}{2\mu}=\frac{\rho}{8}(\vec{z}^{\,+}-\vec{z}^{\,-})^2,\qquad \epsilon^\pm=\frac{\rho}{2}\vec{z}^{\,\pm} \cdot (\vec{z}^{\,\mp}\cdot \nabla \vec{z}^{\,\pm})=\nabla\cdot(\vec{z}^{\,\mp} w^\pm), \label{eq:vdh}
\end{equation}
where the wave energy density $w^\pm$ is given by 
\begin{equation}
	w^\pm=\frac{\rho}{4} (\vec{z}^{\,\pm})^2.\label{eq:energy}
\end{equation}
The second expression for $P_\mathrm{A}$ (relating it directly to \elsasser variables) is not entirely correct for kink waves, because of the additional compressive term in the linearised \elsasser variable (Eq.~\ref{eq:linearelsasser}). Also the right hand side expression for the energy dissipation rate $\epsilon^\pm$ is incorrect for kink waves, because it relies on $\nabla \cdot \vec{z}^{\,\pm}=0$ which is not satisfied by kink waves. Moreover, the expression for the energy density $w^\pm$ has incorrectly included the new compressive term in Eq.~\ref{eq:linearelsasser}. \par
The aim of this work is to calculate the equivalent term of $P_\mathrm{A}$ and $\epsilon^\pm$ for kink waves in a cylindrical loop with a given density contrast. In that way, these expressions may be used in the AWSOM (or equivalent) model to quantify the contribution of uniturbulence \citep{magyar2017} of propagating kink waves or the KHI turbulence from standing kink waves \citep{karampelas2018}. The incorporation of these extra terms forms the basis for the UAWSOM model.

\section{Kink waves using Els\"asser variables}\label{sec:kink}
We consider an equilibrium configuration of a straight cylinder (and associated cylindrical coordinate system $(r,\varphi,z)$) with homogeneous magnetic field $\vec{B}_0=B_0 \vec{e}_z$, without gravity, no background flow and no gas pressure $p=0$. This allows us to take any density profile wanted. We opt for a radial step function in density
\[\rho_0(r)=\begin{cases}
  	\rho_\mathrm{i} & \mbox{for } r\leq R,\\
  	\rho_\mathrm{e} & \mbox{for } r>R,
  \end{cases}
\]
without variation along the other coordinates. The subscripts $i,e$ indicate the interior and exterior region respectively (e.g. $\rho_i$ is the density $\rho_0$ of the interior of the loop). The quantity $R$ represents the radius of the coronal loop. \par
As is well known in the literature \citep[e.g.][]{zaitsev1975,wentzel1979,spruit1981,edwin1983}, the wave solutions for the total pressure perturbation $P'=\frac{B_0b_z}{\mu}$ in this system are described by Bessel functions. For standing waves, we have
\begin{equation}
	P'(r,\varphi,z,t)={\cal R}(r) \cos{\varphi} \cos{(k_z z)} \cos{(\omega t)} , \label{eq:p_stand}
\end{equation}
where the $z$-behaviour is determined by the boundary conditions at the footpoints of the coronal loops that are rooted in the photosphere. For axially propagating waves (upwardly propagating for $k_z>0$ and $\omega>0$, but standing in the $\varphi$-direction)
\begin{equation}
	P'(r,\varphi,z,t)={\cal R}(r) \cos{\varphi} \cos{(k_z z-\omega t)}, \label{eq:p_prop}
\end{equation}
where the radial dependence ${\cal R}(r)$ is given by the continuous function 
\begin{equation}
	{\cal R}(r)= \begin{cases} A\frac{J_1(\kappa_i r)}{J_1(\kappa_i R)} & \mbox{for } r\leq R,\\ A\frac{K_1(\kappa_e r)}{K_1(\kappa_e R)} & \mbox{for } r>R,\end{cases}
\end{equation}
using Bessel function $J_1$, modified Bessel function $K_1$, amplitude $A$, and radial wave number
\begin{equation}
	\kappa^2_\mathrm{i,e}=\left| \frac{\omega^2 - \omega_\mathrm{A}^2}{V_\mathrm{A}^2}\right|,
\end{equation}
with $\omega_\mathrm{A}=k_zV_\mathrm{A}$. The eigenvalue $\omega$ is a root of the dispersion relation:
\begin{equation}
	\frac{\kappa_i}{\rho_i(\omega^2-\omega_\mathrm{Ai}^2)}\frac{J'_1(\kappa_i R)}{J_1(\kappa_i R)}=\frac{\kappa_e}{\rho_e(\omega^2-\omega_\mathrm{Ae}^2)}\frac{K'_1(\kappa_e R)}{K_1(\kappa_e R)},
\end{equation}
in which primes denote derivatives of the Bessel function w.r.t. their arguments.\par

It is not obvious if these solutions obtained for a cold plasma ($v_\mathrm{S}\to 0$) can be applied in the equations (Eq.~\ref{eq:linearelsasser}) for an incompressible plasma ($v_\mathrm{S}\to \infty$). However, the \elsasser formulation also exists for full MHD, as was derived by \citet{marsch1987}. Their Eq.~16-17 show that these full MHD equations have extra terms compared to incompressible MHD. One would expect that these terms are important in the further analysis. However, as for \alfven waves, velocity perturbations of kink waves have no components along the magnetic field in our currently used cold plasma limit, and therefore the linearised contribution of the total pressure in the RHS of Eq.~\ref{eq:fullelsasser} will be zero \citep[following the same reasoning as Eq.~40 in][]{marsch1987}. Therefore, it is at least justified not to keep this term in the following calculations. Moreover, the remaining terms are proportional to the density perturbation $\rho'/\rho_0$ \citep[Eq.~48]{marsch1987}, but these density perturbations linearly scale with the wavenumber  \citep{vd2008} and will go to zero when we take the long-wavelength limit. Furthermore, its dominant contribution will be in determining the longitudinal \elsasser component \citep[first term in Eq.~48 of][while the second term will have a higher order contribution because of the radial gradient]{marsch1987}.

\subsection{Linear quantities}
In order to calculate the \elsasser variables for kink waves, we compute the velocity components $(v_r,v_\varphi,v_z)$ with 
\begin{equation}
	v_r=\frac{1}{\rho_0(\omega^2-\omega_\mathrm{A}^2)} \frac{\partial }{\partial r}\frac{\partial P'}{\partial t}, \quad v_\varphi=\frac{1}{\rho_0(\omega^2-\omega_\mathrm{A}^2)} \frac{1}{r}\frac{\partial }{\partial \varphi}\frac{\partial P'}{\partial t}, \quad v_z=0,
\end{equation}
where the latter statement is true because of the cold plasma limit. The magnetic field components $(b_r,b_\varphi,b_z)$ are calculated with 
\begin{equation}
	\frac{\partial b_r}{\partial t}=B_0\frac{\partial}{\partial z}v_r, \quad \frac{\partial b_\varphi}{\partial t}=B_0\frac{\partial}{\partial z} v_\varphi, \quad b_z=\frac{\mu}{B_0}P'.
\end{equation}
Substituting the following intermediate expressions as derived with Eq.~\ref{eq:p_stand}-\ref{eq:p_prop} \citep[see also][]{yuan2016a}
\begin{align}
	v_r &= \frac{\partial {\cal R}}{\partial r} \frac{\omega}{\rho_0(\omega^2-\omega_\mathrm{A}^2)} \cos{\varphi}\begin{cases}  - \cos{(k_z z)} \sin{(\omega t)} & \mbox{(standing)} \\ \sin{(k_z z-\omega t)} & \mbox{(propagating)} \end{cases} \label{eq:vr}\\
	v_\varphi & = -\frac{{\cal R}}{r} \frac{\omega}{\rho_0(\omega^2-\omega_\mathrm{A}^2)} \sin{\varphi}\begin{cases}  - \cos{(k_z z)} \sin{(\omega t)} & \mbox{(standing)} \\ \sin{(k_z z-\omega t)} & \mbox{(propagating)} \end{cases} \label{eq:vphi}\\
	b_r &= -\frac{\partial {\cal R}}{\partial r} \frac{k_zB_0}{\rho_0(\omega^2-\omega_\mathrm{A}^2)} \cos{\varphi}\begin{cases}  \sin{(k_z z)} \cos{(\omega t)} & \mbox{(standing)} \\ \sin{(k_z z-\omega t)} & \mbox{(propagating)} \end{cases} \\
	b_\varphi &= \frac{{\cal R}}{r} \frac{k_zB_0}{\rho_0(\omega^2-\omega_\mathrm{A}^2)} \sin{\varphi} \begin{cases}  \sin{(k_z z)} \cos{(\omega t)} & \mbox{(standing)} \\ \sin{(k_z z-\omega t)} & \mbox{(propagating)} \end{cases} \\
	b_z & = \frac{\mu {\cal R}}{B_0} \cos{\varphi} \begin{cases} \cos{(k_z z)} \cos{(\omega t)} & \mbox{(standing)} \\ \cos{(k_z z-\omega t)} & \mbox{(propagating)} \end{cases} \label{eq:bz}
\end{align}
in Eq.~\ref{eq:linearelsasser}, it is straightforward to derive the expressions for the \elsasser variables
\begin{align}
	z_r^\pm & = \frac{\partial {\cal R}}{\partial r} \frac{1}{\rho_0(\omega^2-\omega_\mathrm{A}^2)} \cos{\varphi} \begin{cases}  -\omega \cos{(k_z z)} \sin{(\omega t)}\mp \omega_\mathrm{A} \sin{(k_z z)} \cos{(\omega t)} & \mbox{(standing)} \\ (\omega \mp \omega_\mathrm{A}) \sin{(k_z z-\omega t)} & \mbox{(propagating)} \end{cases} \label{eq:zr}\\
	z_\varphi^\pm &= -\frac{{\cal R}}{r} \frac{1}{\rho_0(\omega^2-\omega_\mathrm{A}^2)} \sin{\varphi}\begin{cases}  -\omega \cos{(k_z z)} \sin{(\omega t)}\mp \omega_\mathrm{A} \sin{(k_z z)} \cos{(\omega t)} & \mbox{(standing)} \\ (\omega \mp \omega_\mathrm{A}) \sin{(k_z z-\omega t)} & \mbox{(propagating)} \end{cases} \label{eq:zphi}\\
	z_z^\pm &= \pm \left( \frac{\mu P'}{B_0}\frac{1}{\sqrt{\mu \rho_0}} - V_\mathrm{A}\frac{\rho'}{2\rho_0}\right) = \pm P' \left( \frac{\mu }{B_0}\frac{1}{\sqrt{\mu \rho_0}} - \frac{1}{2\rho_0V_\mathrm{A}}\right) = \pm P' \left(\frac{1}{2\rho_0V_\mathrm{A}}\right) \nonumber \\
	&= \pm \frac{\cal R}{2\rho_0V_\mathrm{A}}\cos{\varphi} \begin{cases} \cos{(k_z z)} \cos{(\omega t)} & \mbox{(standing)} \\ \cos{(k_z z-\omega t)} & \mbox{(propagating)} \end{cases} \label{eq:zz}
\end{align}
Here, and in what follows, we adopt the notation with the curly bracket for the $(z,t)$-dependence, which is the only difference between the standing wave \citep[subject to KH instability][]{terradas2008b, antolin2014} or the propagating waves \citep[subject to uniturbulence][]{magyar2017}. Using this notation, we perform the calculations simultaneously for both cases. Next, we drop the indication (standing/propagating), and we show the expression for the standing wave in the first line, and for the propagating mode in the second line. \par
It is somewhat surprising in Eqs.~\ref{eq:zr}--\ref{eq:zz} that the $z$-component of the \elsasser variable is non-zero, even in the $\beta=0$ limit. This is because the longitudinal magnetic field perturbation is non-zero. The expressions for the perpendicular components of the \elsasser variables of the propagating wave correspond to our earlier intuition: they obey the robust yet elegant relationship that was derived by \citet{magyar2019}
\[ (\omega+\omega_\mathrm{A}) \vec{z}^{\,+}_\perp = (\omega - \omega_\mathrm{A}) \vec{z}^{\,-}_\perp,\]
where $\perp$ are the $(r,\varphi)$-components. However, the expressions for the standing waves do not obey this relationship. This immediately shows the limitation for this relationship, because in its derivation it was assumed that the waves have a $\exp{i(k_zz-\omega t)}$ dependence. This does not hold for the standing waves. In fact, the standing waves are a superposition of two counterpropagating waves. Their $(z,t)$ behaviour has both the upward and downward propagating character, breaking the elegant relationship. This can be made explicit for the $r$ component, for example, using the notation $z_{r,\mathrm{up}}^+$ for the upward propagating waves from before and $z_{r,\mathrm{down}}^+$ for the downward propagating waves with $\cos{(k_zz+\omega t)}$ (assuming both $k_z$ and $\omega$ positive) as $(z,t)$-dependence:
\begin{multline}
	z_{r,\mathrm{standing}}^+=z_{r,\mathrm{up}}^++z_{r,\mathrm{down}}^+=\frac{\partial {\cal R}}{\partial r} \frac{1}{\rho_0(\omega^2-\omega_\mathrm{A}^2)} \cos{\varphi} \left[(\omega - \omega_\mathrm{A}) \sin{(k_z z-\omega t)}-(\omega + \omega_\mathrm{A}) \sin{(k_z z+\omega t)}\right]\\ = \frac{\partial {\cal R}}{\partial r} \frac{1}{\rho_0(\omega^2-\omega_\mathrm{A}^2)} \cos{\varphi} \left[-\omega \cos{(k_zz)}\sin{(\omega t)}-\omega_\mathrm{A}\sin{(k_zz)}\cos{(\omega t)}\right],
\end{multline}
which coincides exactly with the standing wave in Eq.~\ref{eq:zr}. In any case, the expressions for the standing waves are quite surprising, because their $(z,t)$ behaviour is composed of a standing and propagating part! This is made explicit by
\begin{equation}
	-\omega \cos{(k_zz)}\sin{(\omega t)}\mp \omega_\mathrm{A}\sin{(k_zz)}\cos{(\omega t)} = -(\omega \pm \omega_\mathrm{A}) \sin{(k_zz)}\cos{(\omega t)} + \omega \sin{(k_z z-\omega t)}.
\end{equation}
This is reminiscent of the effect of magnetic field twist on standing kink waves, as modelled by \citet{terradas2012,ruderman2015}. There the standing waves also have a partly propagating component in their polarisation. This is because the symmetry is broken by the magnetic twist in their case. Here, the symmetry of the \elsasser variables is broken by the direction of the magnetic field, resulting in a similar propagating part for the standing wave variables. How this is relevant for the development of TWIKH rolls is unclear. 

\subsection{Non-linear quantities}
\subsubsection{Kink wave pressure}
The equivalent of the \alfven wave pressure $P_\mathrm{A}$ (Eq.~\ref{eq:vdh}) for the kink waves $P_\mathrm{k}$ can be computed, using the expression for the magnetic field perturbations:
\begin{multline}
	P_\mathrm{k}=\frac{\vec{b}^2}{2\mu}= \frac{\omega_\mathrm{A}^2}{2\rho_0(\omega^2-\omega_\mathrm{A}^2)^2}\left[ \left(\frac{\partial {\cal R}}{\partial r}\right)^2  \cos^2{\varphi}+\left(\frac{{\cal R}}{r}\right)^2 \sin^2{\varphi}\right] \begin{cases}  \sin^2{(k_z z)} \cos^2{(\omega t)} \\ \sin^2{(k_z z-\omega t)} \end{cases}\\ + \frac{{\cal R}^2}{2\rho_0V_\mathrm{A}^2}\cos^2{\varphi} \begin{cases} \cos^2{(k_z z)} \cos^2{(\omega t)} \\ \cos^2{(k_z z-\omega t)} \end{cases} \label{eq:pk}
\end{multline}
In solar wind modelling, the $z$-component of the gradient of $P_k$ is of interest to provide an extra driving force for the solar wind. This corresponds to the ponderomotive force, as studied by \citet{terradas2004}. While $P_k$ seems to have a constant component from the period averaged $\cos^2$, its effect is cancelled out after taking the gradient. In our simple configuration at least, there is no net force along the magnetic field, but it is rather oscillating with the double wavelength, resulting in oscillating, parallel flows \citep{shestov2017}. Despite there not being a net force in a homogeneous plasma, it will be present when taking the gradient in a WKB system as the derivatives of the coefficients through $\nabla_\parallel B_0$ and $\nabla_\parallel \rho_0$. \\
In the interior part of the loop, the expression (Eq.~\ref{eq:pk}) may be rewritten so that the pressure is proportional to
\begin{multline}P_\mathrm{k}\sim \frac{\omega_\mathrm{A}^2}{2\rho_0(\omega^2-\omega_\mathrm{A}^2)^2}\left[ -\kappa_\mathrm{i}^2J_0(\kappa_\mathrm{i}r)J_2(\kappa_\mathrm{i}r) \cos^2{\varphi}+\frac{J_1^2(\kappa_\mathrm{i}r)}{r^2} \right] \begin{cases}  \sin^2{(k_z z)} \cos^2{(\omega t)} \\ \sin^2{(k_z z-\omega t)} \end{cases}\\ + \frac{J_1^2(\kappa_\mathrm{i}r)}{2\rho_0V_\mathrm{A}^2}\cos^2{\varphi} \begin{cases} \cos^2{(k_z z)} \cos^2{(\omega t)} \\ \cos^2{(k_z z-\omega t)} \end{cases},\end{multline}
with the use of the expressions for the derivatives of Bessel functions: $J_1'=J_0-J_1/z, J_1'=J_2+J_1/z$, and analogously for the exterior part.
\par

\subsubsection{Energy cascade rate}

Let us now calculate the energy cascade rate through $\epsilon^-=\frac{\rho}{2}\vec{z}^{\,-}\cdot (\vec{z}^{\,+}\cdot \nabla \vec{z}^{\,-})$, which expresses the energy dissipation rate for the upward propagating wave. Here we write out the expression $\vec{a}\cdot (\vec{b}\cdot\nabla \vec{a})$ (for a general vector field $\vec{a}$ and $\vec{b}$) in components, 
\begin{align}
	\vec{a}\cdot (\vec{b}\cdot\nabla \vec{a}) &= a_x \left(b_x \frac{\partial}{\partial x} a_x + b_y \frac{\partial}{\partial y} a_x + b_z \frac{\partial}{\partial z} a_x\right)  + a_y \left(b_x \frac{\partial}{\partial x} a_y + b_y \frac{\partial}{\partial y} a_y + b_z \frac{\partial}{\partial z} a_y\right)  + a_z \left(b_x \frac{\partial}{\partial x} a_z + b_y \frac{\partial}{\partial y} a_z + b_z \frac{\partial}{\partial z} a_z\right) \\
	&= b_x \frac{\partial}{\partial x} \frac{a^2}{2} + b_y \frac{\partial}{\partial y} \frac{a^2}{2} + b_z \frac{\partial}{\partial z} \frac{a^2}{2} \\
	&= \vec{b}\cdot \nabla \frac{a^2}{2},
\end{align}
to show that the energy cascade rate $\epsilon^-$ is given in terms of gradients of the wave energy density $w^-$
\begin{equation}
	\epsilon^-=\frac{\rho}{2}\vec{z}^{\,-}\cdot (\vec{z}^{\,+}\cdot \nabla \vec{z}^{\,-})=\vec{z}^{\,+}\cdot \nabla \frac{\rho}{2} \frac{(z^-)^2}{2}=\vec{z}^{\,+}\cdot \nabla w^-, \label{eq:cascaderate}
\end{equation}
if the density $\rho$ is a constant. Here we have defined the wave energy density as 
\begin{equation}
	w^\pm = \frac{\rho (\vec{z}^{\,\pm})^2}{4},
\end{equation}
as was also used in \citet{vanderholst2014} (see Eq.~\ref{eq:energy}). However, in our case, the wave energy density associated with both \elsasser components of the kink wave is split over both $w^{\pm}$. To calculate the full energy in the kink wave, we need to add these. Indeed,
\begin{equation}
	w=w^++w^-=\frac{\rho_0}{4}\sum_{n=0}^{n=1}\left(\vec{v}+ (-1)^n \left( \frac{\vec{b}}{\sqrt{\mu\rho_0}} - \vec{e}_z V_\mathrm{A} \frac{\rho'}{2\rho_0}\right)\right)^2=\rho_0\frac{v^2}{2}+\frac{b^2}{2\mu}+\frac{V_\mathrm{A}^2}{8}\frac{\rho'^2}{\rho_0}.
\end{equation}
As we discussed previously, the latter term is higher order for thin tubes and can be neglected. Then, we find that w is indeed the total energy density associated with the kink wave (propagating or standing).\\ Moreover, the energy cascade rate also needs to be added for both \elsasser variables:
\begin{equation}
	\epsilon=\epsilon^++\epsilon^-,\label{eq:addcascade}
\end{equation}
because this comes from the addition of the differential equation for the evolution of the wave energy \citep[Eq.~22 or 25 in][]{vanderholst2014}.
\par

Using the expressions for the \elsasser variables (Eq.~\ref{eq:zr}--\ref{eq:zz}), we obtain the explicit expression for the wave energy density:
\begin{multline}
	w^\pm= \frac{1}{4}\frac{1}{\rho_0 (\omega^2-\omega_\mathrm{A}^2)^2}\left[\left(\frac{\partial {\cal R}}{\partial r}\right)^2\cos^2{\varphi}+\left(\frac{\cal R}{r}\right)^2\sin^2{\varphi}\right] \begin{cases}  (\omega \cos{(k_z z)} \sin{(\omega t)}\pm \omega_\mathrm{A} \sin{(k_z z)} \cos{(\omega t)})^2 \\ (\omega \mp \omega_\mathrm{A})^2 \sin^2{(k_z z-\omega t)} \end{cases}\\ + \frac{{\cal R}^2 \cos^2{\varphi}}{16 \rho_0 V_\mathrm{A}^2} \begin{cases} \cos^2{(k_z z)} \cos^2{(\omega t)} \\ \cos^2{(k_z z-\omega t)} \end{cases}.
\end{multline}

Now, we have all ingredients to calculate the energy cascade rate. However, inspecting Eq.~\ref{eq:cascaderate} shows that this quantity will have a 3rd power in its harmonic temporal behaviour and a 3rd power in the harmonic $\varphi$ behaviour. A straightforward averaging over a period would reduce the energy cascade rate to 0. Moreover, an averaging over the $\varphi$-direction (i.e. in the cross-section of the cylinder) would reduce it to 0 as well. In modelling of solar wind \alfven turbulence \citep[like][]{vanderholst2014}, this zero average in the time domain is circumvented by approximating $\vec{z}^{\,\pm} \cdot \nabla w^\mp$ by $\sqrt{(\vec{z}^{\,\pm})^2} w^\mp/L$, where $L$ is the correlation length that describes the perpendicular scales of the wave energy distributions. In this step, the time dependence is explicitly left out, so that a non-zero energy cascade rate is obtained. Here, that approach would also be feasible, but when averaging the result over the cylinder cross-section, it would still go to 0, because of the 3rd power of the harmonic $\varphi$ dependence.\par

We thus propose to generalise the method that is used in solar wind \alfven turbulence. Let us take the root-mean-square of the energy cascade rate while averaging over the time direction and the $\varphi$-direction. We thus say that
\begin{equation}
	\langle \epsilon \rangle = \int_0^\infty r dr \left( \int_0^{2\pi} d\varphi \frac{\omega}{2\pi}\int_0^{2\pi/\omega}dt \ \epsilon^2 \right)^{1/2} \label{eq:avenergy}
\end{equation}
is the energy cascade rate averaged over a period and the cross-section of the cylinder. The radial integration is kept outside of the RMS, because otherwise the units of the result would not match ``area$\times$energy/time''.\par

A physical justification for this RMS averaging over time, is that the input waves (kink waves in this case) are cascaded away rapidly enough. By this, we mean that half a cycle will have a positive energy cascade rate, which should in theory be compensated by the negative energy cascade rate in the second half cycle. The idea in the RMS is that the total wave energy is cascaded away to smaller scales in the first half of the cycle, so that it cannot undergo the complete inverse cascade in the second half cycle. Moreover, the modelling of turbulence using only one harmonic is incorrect. It should be seen as an ensemble average over all frequencies, once again justifying the RMS approach, because the frequency and time averaging are a dual process.

\section{Thin-tube limit}\label{sec:thin-tube}
In order to simplify the expressions for the non-linear terms, and to eventually integrate them over the cross-section of the loop, we consider the thin-tube limit. In the thin-tube limit, the frequency of the kink mode is $\omega=\omega_\mathrm{k}=\frac{\rho_i\omega_\mathrm{A,i}^2+\rho_e\omega_\mathrm{A,e}^2}{\rho_i+\rho_e}$, as usual. We use the thin-tube limit as explained by \citet{ruderman2017}. They consider as a small parameter $\delta=R/L=k_zR/\pi\ll 1$, implying quasi-perpendicular propagation. Here we have used $k_z=\pi/L$ in analogy to the fundamental kink mode in a coronal loop, implying that $L$ is half the wavelength. They also use the maximum radial displacement $\eta/R=a$, which is much greater than $\delta$. As a result, they found $V/V_\mathrm{A}=\max{\vec{v}_\perp}/V_\mathrm{A}\sim\delta a$, where we introduce the velocity amplitude $V$.\par
With these assumptions, we find that the radial behaviour simplifies to polynomials and inverse polynomials:
\begin{equation}
	\lim_{\delta\to 0} {\cal R}(r)={\cal T}(r)=\begin{cases} A\frac{r}{R} & \mbox{for } r\leq R\\ A\frac{R}{r} & \mbox{for } r>R\end{cases},
\end{equation}
where we have defined the new function ${\cal T}(r)$. Taking only the radial behaviour of $v_\varphi$ (Eq.~\ref{eq:vphi}) and considering the thin-tube limit, we find that it is 
\begin{align}
	\frac{v_\varphi}{V_\mathrm{A}}&\sim\frac{1}{V_\mathrm{A}}\frac{{\cal T}}{r} \frac{\omega}{\rho_0(\omega^2-\omega_\mathrm{A}^2)}\\
	&=\frac{1}{V_\mathrm{A}\rho_0} \frac{A}{R} \frac{\omega}{(\omega^2-\omega_\mathrm{A}^2)}. \label{eq:amplitudes}
\end{align}
Given that $V/V_\mathrm{A}\sim\delta a$ and $\frac{\omega}{R(\omega^2-\omega_\mathrm{A}^2)}\sim\delta^{-1}$, we must conclude that $\frac{A}{\rho_0V_\mathrm{A}}\sim \delta^2a$. Indeed, in the thin-tube limit, the total pressure and density perturbations are small compared to the velocity perturbations. \par
Taking the thin-tube limit of the expressions in Eqs.~\ref{eq:vr}-\ref{eq:bz}, we obtain the same orders of magnitude as Eq.~6 in \citet{ruderman2017}. These orders of magnitude carry over to the \elsasser variables as well: we find that 
\begin{equation}
	\left| \frac{\vec{z}^{\,\pm}_\perp}{V_\mathrm{A}}\right|\sim \delta a, \qquad \left|\frac{\vec{z}^{\,\pm}_z}{V_\mathrm{A}}\right| \sim \delta^2 a.
\end{equation}
Thus, these kink waves do behave like \alfven waves in the long wavelength limit, in the sense that the \elsasser variables are dominated by the perpendicular components, confirming the results of \citet{goossens2009,goossens2012}. However, the kink waves still keep their velocity component perpendicular to the magnetic surface, in contrast to the \alfven wave \citep{vd2008}. The perpendicular nature of the \elsasser variables is also propagated to the non-linear terms, for which we find (see section~\ref{sec:zgradz})
\begin{equation}
	\left| (\vec{z}^{\,+}\cdot\nabla \vec{z}^{\,-})_\perp\right| \sim (\delta a)^2 \frac{V_\mathrm{A}^2}{R}, \qquad |(\vec{z}^{\,+}\cdot\nabla \vec{z}^{\,-})_z| \sim (\delta a)^2\delta \frac{V_\mathrm{A}^2}{R}.
\end{equation}
It shows that the non-linear generation of parallel flows is an order of magnitude smaller than the turbulence generated in the perpendicular direction.\par

\subsection{Kink wave pressure}
We can calculate the wave pressure $P_\mathrm{k}$ explicitly in the thin-tube limit. First we compute the expression for the pressure in the internal part:
\begin{align}
	P_\mathrm{ki}&= \frac{\omega_\mathrm{A}^2}{2\rho_0(\omega^2-\omega_\mathrm{A}^2)^2}\left[ \left\lbrace \left(\frac{\partial {\cal R}}{\partial r}\right)^2 - \left(\frac{{\cal R}}{r}\right)^2\right\rbrace \cos^2{\varphi}+\left(\frac{{\cal R}}{r}\right)^2 \right] \begin{cases}  \sin^2{(k_z z)} \cos^2{(\omega t)} \\ \sin^2{(k_z z-\omega t)} \end{cases}\nonumber \\ &+ \frac{{\cal R}^2}{2\rho_0V_\mathrm{A}^2}\cos^2{\varphi} \begin{cases} \cos^2{(k_z z)} \cos^2{(\omega t)} \\ \cos^2{(k_z z-\omega t)} \end{cases} \label{eq:pkint} \\ 
	&= \frac{\omega_\mathrm{Ai}^2}{2\rho_\mathrm{i}(\omega^2-\omega_\mathrm{Ai}^2)^2}\left(\frac{A}{R}\right)^2 \begin{cases}  \sin^2{(k_z z)} \cos^2{(\omega t)} \\ \sin^2{(k_z z-\omega t)} \end{cases}+ \frac{A^2r^2}{2\rho_\mathrm{i}R^2V_\mathrm{Ai}^2}\cos^2{\varphi} \begin{cases} \cos^2{(k_z z)} \cos^2{(\omega t)} \\ \cos^2{(k_z z-\omega t)} \end{cases}
\end{align}
Normally, the higher order terms should be taken into account for $\left(\frac{\partial {\cal R}}{\partial r}\right)^2 - \left(\frac{{\cal R}}{r}\right)^2$, but subsequent calculations show that they remain of higher order and do not matter for the end result. Likewise, the second term on the RHS is of order $\delta^4$, and can be ignored compared to the first term.\\
Now we integrate the pressure over the entire cross-section, i.e. from 0 to $R$. 
\begin{equation}
	\iint_\mathrm{i} P_\mathrm{ki} rdrd\varphi = \pi A^2 \frac{\omega_\mathrm{Ai}^2}{2\rho_\mathrm{i}(\omega^2-\omega_\mathrm{Ai}^2)^2} \begin{cases}  \sin^2{(k_z z)} \cos^2{(\omega t)} \\ \sin^2{(k_z z-\omega t)} \end{cases}.
\end{equation}
Here we have omitted the integration constant in the pressure, because it is irrelevant in determining forces.
We thus find that 
\begin{equation}
	\iint_\mathrm{i} P_\mathrm{ki} =V^2 \pi R^2 \frac{\rho_\mathrm{i}\omega_\mathrm{Ai}^2}{2\omega^2} \begin{cases}  \sin^2{(k_z z)} \cos^2{(\omega t)} \\ \sin^2{(k_z z-\omega t)} \end{cases},
\end{equation}
where we have rewritten the expression in terms of the velocity amplitude, with the use of expression Eq.~\ref{eq:amplitudes}. \par
In the exterior part, the expression in the curly brackets in Eq.~\ref{eq:pkint} is also 0 in the leading order (its highest order contribution is $\delta^4\log{\delta}$). We then obtain
\begin{equation}
	P_\mathrm{ke}= \frac{\omega_\mathrm{Ae}^2}{2\rho_\mathrm{e}(\omega^2-\omega_\mathrm{Ae}^2)^2}\left[ \frac{A^2R^2}{r^4} \right] \begin{cases}  \sin^2{(k_z z)} \cos^2{(\omega t)} \\ \sin^2{(k_z z-\omega t)} \end{cases}+ \frac{A^2R^2}{2r^2\rho_\mathrm{e}V_\mathrm{Ae}^2}\cos^2{\varphi} \begin{cases} \cos^2{(k_z z)} \cos^2{(\omega t)} \\ \cos^2{(k_z z-\omega t)} \end{cases}
\end{equation}
We should also ignore the rightmost term, because it is smaller than the ignored term in the first term. Now, we integrate the pressure between $r=R$ and $r=\alpha R$, following the approach in \citet{goossens2013,vd2014,moreels2015}. We then obtain in leading order
\begin{align}
	\iint_\mathrm{e} P_\mathrm{ke}rdrd\varphi &= \frac{\omega_\mathrm{Ae}^2}{2\rho_\mathrm{e}(\omega^2-\omega_\mathrm{Ae}^2)^2}\left[ \pi A^2 \left(\frac{\alpha^2-1}{\alpha^2}\right) \right] \begin{cases}  \sin^2{(k_z z)} \cos^2{(\omega t)} \\ \sin^2{(k_z z-\omega t)} \end{cases} \label{eq:pkext}\\
	&= V^2 \pi R^2 \frac{\rho_\mathrm{e}\omega_\mathrm{Ae}^2}{2\omega^2}\left(\frac{\alpha^2-1}{\alpha^2}\right) \begin{cases}  \sin^2{(k_z z)} \cos^2{(\omega t)} \\ \sin^2{(k_z z-\omega t)} \end{cases}
\end{align}
The other terms are smaller with respect to $\delta$ and have consequently been ignored. However, they contain expressions that have $\log{\alpha}$. So, if $\alpha\to\infty$, we get similar problems as discussed in \citet{moreels2015}. Let us leave them out for now. In leading order, we then find the total pressure integrated over the cross-section
\begin{align}
	\iint P_\mathrm{k}rdrd\varphi & = \iint P_\mathrm{ki}rdrd\varphi + \iint P_\mathrm{ke}rdrd\varphi \\
	& = V^2 \pi R^2\frac{1}{2\omega^2}\left(\rho_\mathrm{i}\omega_\mathrm{Ai}^2+\rho_\mathrm{e}\omega_\mathrm{Ae}^2 \frac{\alpha^2-1}{\alpha^2}\right)\begin{cases}  \sin^2{(k_z z)} \cos^2{(\omega t)} \\ \sin^2{(k_z z-\omega t)} \end{cases}\\
	&= V^2 \pi R^2\left(\frac{\rho_\mathrm{i}+\rho_\mathrm{e}}{2}-\rho_\mathrm{e}\frac{\omega_\mathrm{Ae}^2}{2\omega^2} \frac{1}{\alpha^2}\right)\begin{cases}  \sin^2{(k_z z)} \cos^2{(\omega t)} \\ \sin^2{(k_z z-\omega t)} \end{cases}\\
	&= V^2 \pi R^2\left(\frac{\rho_\mathrm{i}+\rho_\mathrm{e}}{2}-\rho_\mathrm{e}\frac{\omega_\mathrm{Ae}^2}{2\omega^2} f\right)\begin{cases}  \sin^2{(k_z z)} \cos^2{(\omega t)} \\ \sin^2{(k_z z-\omega t)} \end{cases},
\end{align}
where we have used the expression for the long wavelength limit of the kink frequency \begin{equation} \omega_\mathrm{k}^2=\frac{\rho_\mathrm{i}\omega_\mathrm{Ai}^2+\rho_\mathrm{e}\omega_\mathrm{Ae}^2}{\rho_\mathrm{i}+\rho_\mathrm{e}} \label{eq:kink} \end{equation} in the 3rd equality. In the 4th equality, we have connected this to the filling factor $f=1/\alpha^2$, using the same line of thought as \citet{vd2014}. Here the filling factor $f$ is defined as the area of internal cross-sections to external plasma in a multistranded loop \citep[see][for more details]{vd2014}.  \par
This formula shows that quite a few free parameters will be introduced in the UAWSOM model. The density contrast, loop/plume radius and filling factor will play a major role. Parallel gradients of the density, magnetic field $\nabla_\parallel B_0$ and expansion will play a major role in generating a net wave pressure, through their influence on $V$, $R$, densities and filling factor. \par

\subsection{Energy cascade rate}
Given the importance of the energy density $w$, let us have a look at its behaviour in the thin-tube limit. In the internal and external regions of the tube, we find respectively 
\begin{align}
	w^\pm_\mathrm{i} &= \frac{1}{4}\frac{1}{\rho_\mathrm{i} (\omega^2-\omega_\mathrm{Ai}^2)^2}\frac{A^2}{R^2} \begin{cases}  (\omega \cos{(k_z z)} \sin{(\omega t)}\pm \omega_\mathrm{Ai} \sin{(k_z z)} \cos{(\omega t)})^2 \\ (\omega \mp \omega_\mathrm{Ai})^2 \sin^2{(k_z z-\omega t)} \end{cases},\\
	w^\pm_\mathrm{e} &= \frac{1}{4}\frac{1}{\rho_\mathrm{e} (\omega^2-\omega_\mathrm{Ae}^2)^2}\frac{A^2R^2}{r^4} \begin{cases}  (\omega \cos{(k_z z)} \sin{(\omega t)}\pm \omega_\mathrm{Ae} \sin{(k_z z)} \cos{(\omega t)})^2 \\ (\omega \mp \omega_\mathrm{Ae})^2 \sin^2{(k_z z-\omega t)} \end{cases}. \label{eq:wintwext}
\end{align}
This shows that the gradient of the internal energy density $\nabla w^\pm_\mathrm{i}$ only has higher order contributions w.r.t. $\delta$. We do not consider it in the remainder of this subsection. The dominant contribution to the energy cascade rate comes from 
\begin{equation} 
	\epsilon^\mp = z^\pm_{r\mathrm{e}} \frac{\partial}{\partial r} w^\mp_e.
\end{equation}
because these are the only terms that have a contribution at $\delta^3$. We have checked this conclusion in Maple, taking into account the next higher order terms in all quantities, confirming this conclusion. Inserting the appropriate expressions, we find
\begin{equation}
	\epsilon^\mp = \frac{A^3R^3}{r^7} \frac{1}{\rho_\mathrm{e}^2(\omega^2-\omega_\mathrm{Ae}^2)^3} \cos{\varphi} \begin{cases}  (-\omega \cos{(k_z z)} \sin{(\omega t)}\mp \omega_\mathrm{Ae} \sin{(k_z z)} \cos{(\omega t)})(\omega \cos{(k_z z)} \sin{(\omega t)}\mp \omega_\mathrm{Ae} \sin{(k_z z)} \cos{(\omega t)})^2 \\ (\omega \mp \omega_\mathrm{Ae})(\omega \pm \omega_\mathrm{Ae})^2 \sin^3{(k_z z-\omega t)}\end{cases}
\end{equation}
We add the plus and minus component using Eq.~\ref{eq:addcascade} to obtain the total cascade rate:
\begin{equation}
	\epsilon = \frac{A^3R^3}{r^7} \frac{1}{\rho_\mathrm{e}^2(\omega^2-\omega_\mathrm{Ae}^2)^3} \cos{\varphi} \begin{cases}  -2\omega \cos{(k_z z)} \sin{(\omega t)}(\omega^2 \cos^2{(k_z z)} \sin^2{(\omega t)}- \omega_\mathrm{Ae}^2 \sin^2{(k_z z)} \cos^2{(\omega t)}) \\ 2\omega(\omega^2 - \omega_\mathrm{Ae}^2) \sin^3{(k_z z-\omega t)}\end{cases}
\end{equation}

Then, we average this over a cross-section of the cylinder and period, as described in Eq.~\ref{eq:avenergy} but with the external boundary at $\alpha R$ as for the pressure integral (Eq.~\ref{eq:pkext}). We find as result
\begin{align}
	\langle \epsilon \rangle &= \frac{A^3R^3}{\rho_\mathrm{e}^2(\omega^2-\omega_\mathrm{Ae}^2)^3} \left(\frac{1}{5R^5}\frac{\alpha^5-1}{\alpha^5}\right) \left(\sqrt{\pi}\right) \begin{cases}  \frac{1}{2}\vert \omega \cos{(k_z z)}\vert \sqrt{ 4\omega^4 \cos^4{(k_z z)}+(\omega^2 \cos^2{(k_z z)}- \omega_\mathrm{Ae}^2 \sin^2{(k_z z)})^2} \\ 2\vert\omega(\omega^2 - \omega_\mathrm{Ae}^2)\vert \sqrt{\frac{5}{16}} \end{cases}\\
	& = V^3 \frac{\sqrt{\pi} R}{10} \frac{\alpha^5-1}{\alpha^5} \frac{\rho_\mathrm{e}}{\omega^3} \begin{cases}  \vert \omega \cos{(k_z z)}\vert \sqrt{ 4\omega^4 \cos^4{(k_z z)}+(\omega^2 \cos^2{(k_z z)}- \omega_\mathrm{Ae}^2 \sin^2{(k_z z)})^2} \\ \sqrt{5}\vert\omega(\omega^2 - \omega_\mathrm{Ae}^2)\vert \end{cases} \label{eq:energycascade}
\end{align}
As with the wave pressure, this may be related to the filling factor $f=1/\alpha^2$ by
\begin{equation}
	\langle \epsilon \rangle= V^3 \frac{\sqrt{\pi} R}{10} \left(1-f^{5/2}\right) \frac{\rho_\mathrm{e}}{\omega^3} \begin{cases}  \vert \omega \cos{(k_z z)}\vert \sqrt{ 4\omega^4 \cos^4{(k_z z)}+(\omega^2 \cos^2{(k_z z)}- \omega_\mathrm{Ae}^2 \sin^2{(k_z z)})^2} \\ \sqrt{5}\vert\omega(\omega^2 - \omega_\mathrm{Ae}^2)\vert \end{cases},
\end{equation}
which is the final expression needed for the UAWSOM model. 

\subsection{Damping in uniturbulence}
\begin{figure}
	\plotone{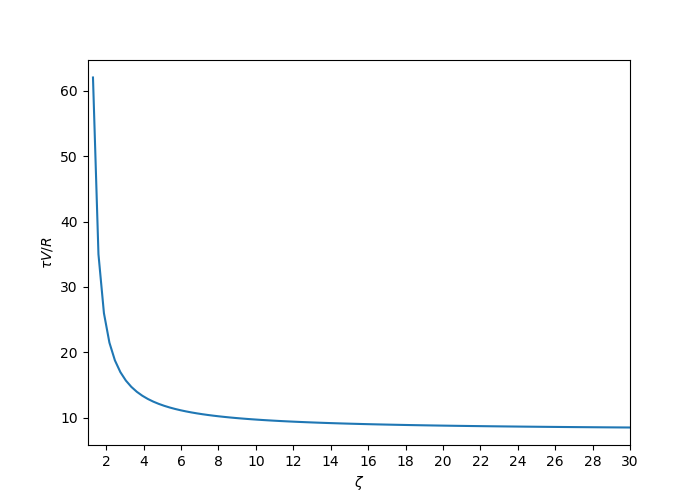}
	\caption{$\tau V/R$ for a propagating kink wave as a function of the density contrast $\zeta$.}
	\label{fig:tau}
\end{figure}

Let us now concentrate on the energy cascade rate for uniturbulence, i.e. we consider only the propagating waves. We have that the energy density averaged over a cross-section is (Eq.~\ref{eq:wintwext} with integration boundary $\alpha\to\infty$)
\begin{equation}
	\langle w \rangle = \pi R^2 \frac{\rho_\mathrm{i}+\rho_\mathrm{e}}{2}V^2,
\end{equation}
which coincides with the expression of  \citet{goossens2013}.
Taking the ratio of $\langle w \rangle/\langle \epsilon \rangle$, we obtain the time scale for the energy cascade $\tau$:
\begin{equation}
	\tau=\frac{\langle w \rangle}{\langle \epsilon \rangle}=\sqrt{5\pi} \frac{R}{V} \frac{\rho_\mathrm{i}+\rho_\mathrm{e}}{\rho_\mathrm{e}}\frac{\omega^2 }{\vert\omega^2 - \omega_\mathrm{Ae}^2\vert }.
\end{equation}
This time scale is apparently mainly determined by the radius and velocity amplitude. Rewriting it in terms of the density contrast $\zeta=\rho_\mathrm{i}/\rho_\mathrm{e}$ yields
\begin{equation}
	\tau=\sqrt{5\pi} \frac{R}{V} \frac{2 (\zeta+1)}{\vert \zeta-1\vert }. \label{eq:damping}
\end{equation}
When $\zeta\to 1$, the damping time goes to infinity, implying that there is no damping in the system. A density contrast is crucial for uniturbulence, as was also discussed by \citet{magyar2019}. When $\zeta \to \infty$, the damping saturates and is minimal at $\tau\to 2\sqrt{5\pi}\frac{R}{V}$. The general shape for $\tau V/R$ as a function of the density contrast $\zeta$ is shown in Fig.~\ref{fig:tau}.
As an example, let us consider a velocity amplitude of $V=22\mathrm{km/s}$ and radius $R=250\mathrm{km}$ for a coronal plume \citep[as was used in][]{pant2019}: then $R/V=11\mathrm{s}$. For a density contrast $\rho_\mathrm{i}/\rho_\mathrm{e}=3$, we find a timescale $\tau\sim 180\mathrm{s}$, and for a density contrast of $\zeta=10$, we have $\tau\sim 110\mathrm{s}$. These times are relatively short compared to their driving period of around 400s, and it explains why these simulations develop uniturbulence so quickly. \\
As another example, we could think about larger scale coronal plumes with a radius of $R=1\mathrm{Mm}$, and a low velocity amplitude of $V=4\mathrm{km/s}$. Then $R/V=250\mathrm{s}$, with a time scale of $\tau\sim 3960\mathrm{s}$ for a density contrast of $\zeta=3$. In this second example, the energy cascade time is long compared to a driving period of 300s. \par
The expression for the energy cascade time $\tau$ could be even further simplified using $V=\omega \eta$ and our earlier notation $\eta=aR$. Then the factor $R/V$ would become $R/\omega\eta=1/\omega a=P/2\pi a$ (for a period $P=2\pi/\omega$), in which $a$ is the oscillation amplitude in plume radii, finally arriving at 
\begin{equation}
	\tau=\sqrt{5\pi} \frac{P}{2\pi a} \frac{2 (\zeta+1)}{\vert \zeta-1\vert }.\label{eq:tau}
\end{equation}

\section{Conclusions}
We have reformulated the well-known kink mode solution for transverse waves in a uniform, field-aligned cylinder into the \elsasser formalism. We have done this for both standing waves (that are subject to KHI) and for propagating waves (subject to uniturbulence). Using the expressions for the \elsasser variables, we have computed the kink wave pressure and the energy cascade rate. These expressions are to be added to the \alfven wave turbulence equivalent expressions in the AWSOM model \citep{vanderholst2014}, in order to formulate the UAWSOM model. In this process, care needs to be taken in the formulation: our energy cascade rate works for the sum of the two \elsasser components of the upward travelling kink waves (let's call them $\vec{z}^{\,\pm}_\mathrm{up}$). However, the AWSOM model also tracks the downward propagating waves, which are characterised by $\vec{z}^{\,\pm}_\mathrm{down}$. For these reflected waves, the damping by uniturbulence is probably not important. \par
Especially the expression for the uniturbulence damping is of interest. Eq.~\ref{eq:tau} predicts that the damping time of the wave $\tau/P$ is inversely proportional to the wave amplitude normalised to the plume radius $a=\mathrm{max}(\vec{\xi})/R$. For realistic parameters of previous simulations and observations, we find damping times of 100-4000s, showing that uniturbulent damping could play an important role in the open field regions where the fast solar wind is generated. Given its dependence on the density contrast, uniturbulence is probably more important in the structured lower part of the solar corona and coronal hole. \par
It is important that our theoretical results are verified against direct numerical simulations of turbulence formation. Direct numerical simulations of turbulence were already done in \citet{andres2017} for homogeneous plasmas, and needs to be modified to structured plasmas (considering both compressible and incompressible MHD). A first start to this was done in \citet{magyar2019a}, but this will be extended in future work. \par
Our formulation does not clarify the turbulent properties of uniturbulence. It only shows the energy input at large scales, and does not show the evolution of the small scales and high frequencies. Our formulation does not contain information about the power law index in a structured medium. However, our results might be related to the $k^{-1}$ power law that becomes more dominant in the inner heliosphere. \par

\acknowledgments
TVD was supported by the European Research Council (ERC) under the European Union's Horizon 2020 research and innovation programme (grant agreement No 724326) and the C1 grant TRACEspace of Internal Funds KU Leuven.  BL was supported by the National Natural Science Foundation of China (41674172, 11761141002). The research benefitted greatly from discussions at ISSI-BJ.

%






\appendix

\section{Expressions for the non-linear driving terms}\label{sec:zgradz}

For completeness, here we give the expressions for the non-linear contributions of the \elsasser variables  $\vec{z}^{\,\mp}\cdot\nabla \vec{z}^{\,\pm}$ to Eq.~\ref{eq:linearmhd}. This will give information on the non-linear generation of higher $m$ modes, and could be the start of a description of the temporal evolution of uniturbulence or KHI for the (propagating or standing) kink mode. Using the earlier intermediate results (Eq.~\ref{eq:zr}--\ref{eq:zz}), we obtain:
\begin{align}
	(\vec{z}^{\,\mp}\cdot\nabla \vec{z}^{\,\pm})_r &= \frac{1}{\rho_0^2(\omega^2-\omega_\mathrm{A}^2)^2}\left[\frac{\partial {\cal R}}{\partial r}\frac{\partial^2 {\cal R}}{\partial r^2}\cos^2{\varphi}+\frac{{\cal R}}{r^2}\left(\frac{\partial {\cal R}}{\partial r}-\frac{\cal R}{r}\right)\sin^2{\varphi}\right]\begin{cases} \omega^2\cos^2{(k_z z)}\sin^2{(\omega t)}-\omega_\mathrm{A}^2\sin^2{(k_z z)}\cos^2{(\omega t)}\\ (\omega^2-\omega_\mathrm{A}^2)\sin^2{(k_zz -\omega t)}\end{cases} \nonumber \\ & \mp \frac{k_z}{2V_\mathrm{A}\rho_0^2(\omega^2-\omega_\mathrm{A}^2)}{\cal R}\frac{\partial {\cal R}}{\partial r}\cos^2{\varphi} \begin{cases} (\omega \sin{(k_z z)}\sin{(\omega t)}\mp\omega_\mathrm{A}\cos{(k_z z)}\cos{(\omega t)})\cos{(k_z z)}\cos{(\omega t)}\\ (\omega\mp\omega_\mathrm{A})\cos^2{(k_zz-\omega t)}\end{cases} \label{eq:zzr} \\
	(\vec{z}^{\,\mp}\cdot\nabla \vec{z}^{\,\pm})_\varphi &= \frac{1}{\rho_0^2(\omega^2-\omega_\mathrm{A}^2)^2}\left[-\frac{1}{r}\left(\frac{\partial {\cal R}}{\partial r}\right)^2+\frac{{\cal R}^2}{r^3}\right]\cos{\varphi}\sin{\varphi}\begin{cases} \omega^2\cos^2{(k_z z)}\sin^2{(\omega t)}-\omega_\mathrm{A}^2\sin^2{(k_z z)}\cos^2{(\omega t)}\\ (\omega^2-\omega_\mathrm{A}^2)\sin^2{(k_zz -\omega t)}\end{cases} \nonumber \\ & \pm \frac{k_z}{2V_\mathrm{A}\rho_0^2(\omega^2-\omega_\mathrm{A}^2)}\frac{{\cal R}^2}{r}\cos{\varphi}\sin{\varphi} \begin{cases} (\omega \sin{(k_z z)}\sin{(\omega t)}\mp\omega_\mathrm{A}\cos{(k_z z)}\cos{(\omega t)})\cos{(k_z z)}\cos{(\omega t)}\\ (\omega\mp\omega_\mathrm{A})\cos^2{(k_zz-\omega t)}\end{cases} \label{eq:zzphi} \\
	(\vec{z}^{\,\mp}\cdot\nabla \vec{z}^{\,\pm})_z &= \mp \frac{1}{2V_\mathrm{A}\rho_0^2(\omega^2-\omega_\mathrm{A}^2)} \left[\left(\frac{\partial {\cal R}}{\partial r}\right)^2\cos^2{\varphi}+\frac{{\cal R}^2}{r^2}\sin^2{\varphi}\right]\begin{cases} (\omega \cos{(k_z z)}\sin{(\omega t)}\mp\omega_\mathrm{A}\sin{(k_z z)}\cos{(\omega t)})\cos{(k_z z)}\cos{(\omega t)} \\ -(\omega\pm\omega_\mathrm{A})\cos{(k_zz-\omega t)}\sin{(k_zz-\omega t)}\end{cases} \nonumber \\ & + \frac{k_z}{\rho_0^2V_\mathrm{A}^2}{\cal R}^2\cos^2{\varphi} \begin{cases} \cos{(k_z z)}\sin{(k_z z)}\cos^2{(\omega t)} \\ \cos{(k_zz-\omega t)}\sin{(k_zz-\omega t)}\end{cases} \label{eq:zzz}
\end{align}

For the propagating kink waves, the maximum of the non-linear terms along $z$ is in the same location as the velocity peaks. For the standing waves, the maximum of the non-linear terms has contributions both at the loop top and at the footpoint, and indeed this gives quite some numerical instability at the footpoint boundary. This agrees well with the models of \citet{karampelas2019}, where the instability is found in the velocity near the loop top but in the currents near the footpoint. \\
In accordance with our earlier result that the ponderomotive force is acting along the magnetic field, we find that for the standing wave there is a non-linear contribution of the $\vec{z}^{\,+}\cdot\nabla\vec{z}^{\,-}$ in the $z$-direction with a non-zero period average. This leads to the generation of flows. 
\par

We can also expand the expressions for the non-linear generation of waves (Eqs.~\ref{eq:zzr}-\ref{eq:zzz}). However, many of the leading order terms are 0. We therefore need the higher order expansion of ${\cal R}(r)$, which reads
\begin{equation}
	\lim_{\delta\to 0} {\cal R}(r)=\begin{cases} A\frac{r}{R}\left(1-\frac{(\kappa_\mathrm{i}r)^2}{8}\right) & \mbox{for } r\leq R\\ A\frac{R}{r} +\frac{1}{2}AR\kappa_\mathrm{e}^2 r\log{\kappa_\mathrm{e}r} & \mbox{for } r>R\end{cases}.
\end{equation}
Then we obtain the following expressions for the leading terms (in $\delta$) in the non-linear forces in the interior region:
\begin{align}
	(\vec{z}^{\,\mp}\cdot\nabla \vec{z}^{\,\pm})_r &= \frac{A^2}{R^2}r \left(-\frac{1}{\rho_0^2(\omega^2-\omega_\mathrm{A}^2)^2}\frac{\kappa_\mathrm{i}^2}{4} \left[1+2\cos^2{\varphi}\right]\begin{cases} \omega^2\cos^2{(k_z z)}\sin^2{(\omega t)}-\omega_\mathrm{A}^2\sin^2{(k_z z)}\cos^2{(\omega t)}\\ (\omega^2-\omega_\mathrm{A}^2)\sin^2{(k_zz -\omega t)}\end{cases}\right. \nonumber \\ & \left. \mp \frac{k_z}{2V_\mathrm{A}\rho_0^2(\omega^2-\omega_\mathrm{A}^2)}\cos^2{\varphi} \begin{cases} (\omega \sin{(k_z z)}\sin{(\omega t)}\mp\omega_\mathrm{A}\cos{(k_z z)}\cos{(\omega t)})\cos{(k_z z)}\cos{(\omega t)}\\ (\omega\mp\omega_\mathrm{A})\cos^2{(k_zz-\omega t)}\end{cases}\right) \label{eq:zzr-lin-int} \\
	(\vec{z}^{\,\mp}\cdot\nabla \vec{z}^{\,\pm})_\varphi &= \frac{A^2}{R^2}r \cos{\varphi}\sin{\varphi} \left(\frac{1}{\rho_0^2(\omega^2-\omega_\mathrm{A}^2)^2}\frac{\kappa_\mathrm{i}^2}{2}\begin{cases} \omega^2\cos^2{(k_z z)}\sin^2{(\omega t)}-\omega_\mathrm{A}^2\sin^2{(k_z z)}\cos^2{(\omega t)}\\ (\omega^2-\omega_\mathrm{A}^2)\sin^2{(k_zz -\omega t)}\end{cases}\right. \nonumber \\ & \left. \pm \frac{k_z}{2V_\mathrm{A}\rho_0^2(\omega^2-\omega_\mathrm{A}^2)} \begin{cases} (\omega \sin{(k_z z)}\sin{(\omega t)}\mp\omega_\mathrm{A}\cos{(k_z z)}\cos{(\omega t)})\cos{(k_z z)}\cos{(\omega t)}\\ (\omega\mp\omega_\mathrm{A})\cos^2{(k_zz-\omega t)}\end{cases}\right) \label{eq:zzphi-lin-int} \\
	(\vec{z}^{\,\mp}\cdot\nabla \vec{z}^{\,\pm})_z &= \frac{A^2}{R^2}\left( \mp \frac{1}{2V_\mathrm{A}\rho_0^2(\omega^2-\omega_\mathrm{A}^2)} \begin{cases} (\omega \cos{(k_z z)}\sin{(\omega t)}\mp\omega_\mathrm{A}\sin{(k_z z)}\cos{(\omega t)})\cos{(k_z z)}\cos{(\omega t)} \\ -(\omega\pm\omega_\mathrm{A})\cos{(k_zz-\omega t)}\sin{(k_zz-\omega t)}\end{cases}\right. \nonumber \\ & \left. + \frac{k_z r^2 \cos^2{\varphi}}{\rho_0^2V_\mathrm{A}^2} \begin{cases} \cos{(k_z z)}\sin{(k_z z)}\cos^2{(\omega t)} \\ \cos{(k_zz-\omega t)}\sin{(k_zz-\omega t)}\end{cases}\right) \label{eq:zzz-lin-int}
\end{align}
From these expressions, it becomes apparent that the perpendicular components of $(\vec{z}^\mp\cdot\nabla\vec{z}^\pm)_\perp$ are peaking on the tube boundary. This then confirms the simulation results of \citet{terradas2008b} that show that the boundary becomes unstable to KHI first. Also, it agrees with the results of \citet{magyar2019} showing that uniturbulence develops first at the interface.\\
For the perpendicular components $r,\varphi$, the factor in the round brackets is of order ${\cal O}(\delta^0)$. We see that the perpendicular components in the non-linear term are linearly increasing towards the edge of the loop, but since $A\sim {\cal O}(\delta^2)$ (see Sec.~\ref{sec:thin-tube}) they are of the order ${\cal O}(\delta^4)$. This agrees with the conclusions of \citet{ruderman2017} who shows that there is no non-linear solution in the interior of the tube. The non-linear term in the interior of the tube is dominated by the first term of the $z$-component: it is of the order ${\cal O}(\delta^3)$. It is constant in the cross-section. \par

In the exterior region, we obtain the following expressions:
\begin{align}
	(\vec{z}^{\,\mp}\cdot\nabla \vec{z}^{\,\pm})_r &= -2\frac{A^2R^2}{r^5} \frac{1}{\rho_0^2(\omega^2-\omega_\mathrm{A}^2)^2} \begin{cases} \omega^2\cos^2{(k_z z)}\sin^2{(\omega t)}-\omega_\mathrm{A}^2\sin^2{(k_z z)}\cos^2{(\omega t)}\\ (\omega^2-\omega_\mathrm{A}^2)\sin^2{(k_zz -\omega t)}\end{cases} \nonumber \\ & +{\cal O}\left(\frac{\delta^4}{r}\right) \label{eq:zzr-lin-ext} \\
	(\vec{z}^{\,\mp}\cdot\nabla \vec{z}^{\,\pm})_\varphi &= \frac{A^2R^2}{r^3}\log{\kappa_\mathrm{i}r} \cos{\varphi}\sin{\varphi} \frac{1}{\rho_0^2(\omega^2-\omega_\mathrm{A}^2)^2}2\kappa_\mathrm{i}^2\begin{cases} \omega^2\cos^2{(k_z z)}\sin^2{(\omega t)}-\omega_\mathrm{A}^2\sin^2{(k_z z)}\cos^2{(\omega t)}\\ (\omega^2-\omega_\mathrm{A}^2)\sin^2{(k_zz -\omega t)}\end{cases} \nonumber \\ & +{\cal O}\left(\frac{\delta^4}{r^3}\right) \label{eq:zzphi-lin-ext} \\
	(\vec{z}^{\,\mp}\cdot\nabla \vec{z}^{\,\pm})_z &= \mp \frac{A^2R^2}{r^4}\frac{1}{2V_\mathrm{A}\rho_0^2(\omega^2-\omega_\mathrm{A}^2)} \begin{cases} (\omega \cos{(k_z z)}\sin{(\omega t)}\mp\omega_\mathrm{A}\sin{(k_z z)}\cos{(\omega t)})\cos{(k_z z)}\cos{(\omega t)} \\ -(\omega\pm\omega_\mathrm{A})\cos{(k_zz-\omega t)}\sin{(k_zz-\omega t)}\end{cases} \nonumber \\ & +{\cal O}\left(\frac{\delta^5}{r^2}\right) \label{eq:zzz-lin-ext}
\end{align}
The leading term of the $\varphi$-component is $\delta^4\log{\delta}$ in the exterior, while the $z$-component remains $\delta^3$. In the $z$-component, we ignored a term that may become infinite when averaging the quantity until $\alpha R$ and $\alpha \to \infty$. The $r$-component's leading term is $\delta^2$ and is the dominant contribution to the non-linear term in the entire domain, as was also found by \citet{ruderman2017}. It is strange that it does not contain a $\varphi$-dependence, which would indicate the cascade to higher $m$.


\bibliography{../../refs/refs}

\begin{thebibliography}{}
\expandafter\ifx\csname natexlab\endcsname\relax\def\natexlab#1{#1}\fi
\providecommand{\url}[1]{\href{#1}{#1}}
\providecommand{\dodoi}[1]{doi:~\href{http://doi.org/#1}{\nolinkurl{#1}}}
\providecommand{\doeprint}[1]{\href{http://ascl.net/#1}{\nolinkurl{http://ascl.net/#1}}}
\providecommand{\doarXiv}[1]{\href{https://arxiv.org/abs/#1}{\nolinkurl{https://arxiv.org/abs/#1}}}

\bibitem[{{Andr{\'e}s} {et~al.}(2017){Andr{\'e}s}, {Clark di Leoni}, {Mininni},
  {Dmitruk}, {Sahraoui}, \& {Matthaeus}}]{andres2017}
{Andr{\'e}s}, N., {Clark di Leoni}, P., {Mininni}, P.~D., {et~al.} 2017,
  Physics of Plasmas, 24, 102314, \dodoi{10.1063/1.4997990}

\bibitem[{{Anfinogentov} {et~al.}(2015){Anfinogentov}, {Nakariakov}, \&
  {Nistic{\`o}}}]{anfinogentov2015}
{Anfinogentov}, S.~A., {Nakariakov}, V.~M., \& {Nistic{\`o}}, G. 2015, \aap,
  583, A136, \dodoi{10.1051/0004-6361/201526195}

\bibitem[{{Antolin} {et~al.}(2008){Antolin}, {Shibata}, {Kudoh}, {Shiota}, \&
  {Brooks}}]{antolin2008}
{Antolin}, P., {Shibata}, K., {Kudoh}, T., {Shiota}, D., \& {Brooks}, D. 2008,
  \apj, 688, 669, \dodoi{10.1086/591998}

\bibitem[{{Antolin} {et~al.}(2014){Antolin}, {Yokoyama}, \& {Van
  Doorsselaere}}]{antolin2014}
{Antolin}, P., {Yokoyama}, T., \& {Van Doorsselaere}, T. 2014, \apjl, 787, L22,
  \dodoi{10.1088/2041-8205/787/2/L22}

\bibitem[{Bruno \& Carbone(2005)}]{bruno2005}
Bruno, R., \& Carbone, V. 2005, Living Reviews in Solar Physics, 2, 4

\bibitem[{{Buchlin} \& {Velli}(2007)}]{buchlin2007}
{Buchlin}, E., \& {Velli}, M. 2007, \apj, 662, 701, \dodoi{10.1086/512765}

\bibitem[{{Chandran} \& {Hollweg}(2009)}]{chandran2009}
{Chandran}, B. D.~G., \& {Hollweg}, J.~V. 2009, \apj, 707, 1659,
  \dodoi{10.1088/0004-637X/707/2/1659}

\bibitem[{{Chandran} \& {Perez}(2019)}]{chandran2019}
{Chandran}, B. D.~G., \& {Perez}, J.~C. 2019, Journal of Plasma Physics, 85,
  905850409, \dodoi{10.1017/S0022377819000540}

\bibitem[{{De Pontieu} {et~al.}(2007){De Pontieu}, {McIntosh}, {Carlsson},
  {Hansteen}, {Tarbell}, {Schrijver}, {Title}, {Shine}, {Tsuneta}, {Katsukawa},
  {Ichimoto}, {Suematsu}, {Shimizu}, \& {Nagata}}]{depontieu2007}
{De Pontieu}, B., {McIntosh}, S.~W., {Carlsson}, M., {et~al.} 2007, Science,
  318, 1574, \dodoi{10.1126/science.1151747}

\bibitem[{{Dmitruk} {et~al.}(2002){Dmitruk}, {Matthaeus}, {Milano}, {Oughton},
  {Zank}, \& {Mullan}}]{dmitruk2002}
{Dmitruk}, P., {Matthaeus}, W.~H., {Milano}, L.~J., {et~al.} 2002, \apj, 575,
  571, \dodoi{10.1086/341188}

\bibitem[{{Edwin} \& {Roberts}(1983)}]{edwin1983}
{Edwin}, P.~M., \& {Roberts}, B. 1983, \solphys, 88, 179

\bibitem[{{Goddard} {et~al.}(2016){Goddard}, {Nistic{\`o}}, {Nakariakov}, \&
  {Zimovets}}]{goddard2016}
{Goddard}, C.~R., {Nistic{\`o}}, G., {Nakariakov}, V.~M., \& {Zimovets}, I.~V.
  2016, \aap, 585, A137, \dodoi{10.1051/0004-6361/201527341}

\bibitem[{{Goldreich} \& {Sridhar}(1995)}]{goldreich1995}
{Goldreich}, P., \& {Sridhar}, S. 1995, \apj, 438, 763, \dodoi{10.1086/175121}

\bibitem[{{Goossens} {et~al.}(2012){Goossens}, {Andries}, {Soler}, {Van
  Doorsselaere}, {Arregui}, \& {Terradas}}]{goossens2012}
{Goossens}, M., {Andries}, J., {Soler}, R., {et~al.} 2012, \apj, 753, 111,
  \dodoi{10.1088/0004-637X/753/2/111}

\bibitem[{{Goossens} {et~al.}(2009){Goossens}, {Terradas}, {Andries},
  {Arregui}, \& {Ballester}}]{goossens2009}
{Goossens}, M., {Terradas}, J., {Andries}, J., {Arregui}, I., \& {Ballester},
  J.~L. 2009, \aap, 503, 213, \dodoi{10.1051/0004-6361/200912399}

\bibitem[{{Goossens} {et~al.}(2013){Goossens}, {Van Doorsselaere}, {Soler}, \&
  {Verth}}]{goossens2013}
{Goossens}, M., {Van Doorsselaere}, T., {Soler}, R., \& {Verth}, G. 2013, \apj,
  768, 191, \dodoi{10.1088/0004-637X/768/2/191}

\bibitem[{{Heyvaerts} \& {Priest}(1983)}]{heyvaerts1983}
{Heyvaerts}, J., \& {Priest}, E.~R. 1983, \aap, 117, 220

\bibitem[{{Karampelas} \& {Van Doorsselaere}(2018)}]{karampelas2018}
{Karampelas}, K., \& {Van Doorsselaere}, T. 2018, \aap, 610, L9,
  \dodoi{10.1051/0004-6361/201731646}

\bibitem[{{Karampelas} {et~al.}(2019){Karampelas}, {Van Doorsselaere}, \&
  {Guo}}]{karampelas2019}
{Karampelas}, K., {Van Doorsselaere}, T., \& {Guo}, M. 2019, \aap, 623, A53,
  \dodoi{10.1051/0004-6361/201834309}

\bibitem[{Kolmogorov(1962)}]{kolmogorov1962}
Kolmogorov, A.~N. 1962, Journal of Fluid Mechanics, 13, 82–85,
  \dodoi{10.1017/S0022112062000518}

\bibitem[{{Magyar} \& {Van Doorsselaere}(2016)}]{magyar2016b}
{Magyar}, N., \& {Van Doorsselaere}, T. 2016, \aap, 595, A81,
  \dodoi{10.1051/0004-6361/201629010}

\bibitem[{Magyar {et~al.}({2017})Magyar, Van~Doorsselaere, \&
  Goossens}]{magyar2017}
Magyar, N., Van~Doorsselaere, T., \& Goossens, M. {2017}, {Nat. Sci. Rep.},
  {7}, \dodoi{{10.1038/s41598-017-13660-1}}

\bibitem[{{Magyar} {et~al.}(2019{\natexlab{a}}){Magyar}, {Van Doorsselaere}, \&
  {Goossens}}]{magyar2019}
{Magyar}, N., {Van Doorsselaere}, T., \& {Goossens}, M. 2019{\natexlab{a}},
  \apj, 882, 50, \dodoi{10.3847/1538-4357/ab357c}

\bibitem[{{Magyar} {et~al.}(2019{\natexlab{b}}){Magyar}, {Van Doorsselaere}, \&
  {Goossens}}]{magyar2019a}
---. 2019{\natexlab{b}}, \apj, 873, 56, \dodoi{10.3847/1538-4357/ab04a7}

\bibitem[{{Marsch} \& {Mangeney}(1987)}]{marsch1987}
{Marsch}, E., \& {Mangeney}, A. 1987, \jgr, 92, 7363,
  \dodoi{10.1029/JA092iA07p07363}

\bibitem[{{Matthaeus} {et~al.}(1999){Matthaeus}, {Zank}, {Oughton}, {Mullan},
  \& {Dmitruk}}]{matthaeus1999}
{Matthaeus}, W.~H., {Zank}, G.~P., {Oughton}, S., {Mullan}, D.~J., \&
  {Dmitruk}, P. 1999, \apjl, 523, L93, \dodoi{10.1086/312259}

\bibitem[{{McIntosh} \& {De Pontieu}(2012)}]{mcintosh2012}
{McIntosh}, S.~W., \& {De Pontieu}, B. 2012, \apj, 761, 138,
  \dodoi{10.1088/0004-637X/761/2/138}

\bibitem[{{McIntosh} {et~al.}(2011){McIntosh}, {de Pontieu}, {Carlsson},
  {Hansteen}, {Boerner}, \& {Goossens}}]{mcintosh2011}
{McIntosh}, S.~W., {de Pontieu}, B., {Carlsson}, M., {et~al.} 2011, \nat, 475,
  477, \dodoi{10.1038/nature10235}

\bibitem[{{Miki{\'c}} {et~al.}(2018){Miki{\'c}}, {}, {Downs}, {Linker},
  {Caplan}, {Mackay}, {Upton}, {Riley}, {Lionello}, {T{\"o}r{\"o}k}, {Titov},
  {Wijaya}, {Druckm{\"u}ller}, {Pasachoff}, \& {Carlos}}]{mikic2018}
{Miki{\'c}}, {}, Z., {Downs}, C., {et~al.} 2018, Nature Astronomy, 2, 913,
  \dodoi{10.1038/s41550-018-0562-5}

\bibitem[{{Moreels} {et~al.}(2015){Moreels}, {Van Doorsselaere}, {Grant},
  {Jess}, \& {Goossens}}]{moreels2015}
{Moreels}, M.~G., {Van Doorsselaere}, T., {Grant}, S.~D.~T., {Jess}, D.~B., \&
  {Goossens}, M. 2015, \aap, 578, A60, \dodoi{10.1051/0004-6361/201425468}

\bibitem[{{Moriyasu} {et~al.}(2004){Moriyasu}, {Kudoh}, {Yokoyama}, \&
  {Shibata}}]{moriyasu2004}
{Moriyasu}, S., {Kudoh}, T., {Yokoyama}, T., \& {Shibata}, K. 2004, \apj, 601,
  L107

\bibitem[{{Nechaeva} {et~al.}(2019){Nechaeva}, {Zimovets}, {Nakariakov}, \&
  {Goddard}}]{nechaeva2019}
{Nechaeva}, A., {Zimovets}, I.~V., {Nakariakov}, V.~M., \& {Goddard}, C.~R.
  2019, \apjs, 241, 31, \dodoi{10.3847/1538-4365/ab0e86}

\bibitem[{{Nistic{\`o}} {et~al.}(2013){Nistic{\`o}}, {Nakariakov}, \&
  {Verwichte}}]{nistico2013}
{Nistic{\`o}}, G., {Nakariakov}, V.~M., \& {Verwichte}, E. 2013, \aap, 552,
  A57, \dodoi{10.1051/0004-6361/201220676}

\bibitem[{{Pant} {et~al.}(2019){Pant}, {Magyar}, {Van Doorsselaere}, \&
  {Morton}}]{pant2019}
{Pant}, V., {Magyar}, N., {Van Doorsselaere}, T., \& {Morton}, R.~J. 2019,
  \apj, 881, 95, \dodoi{10.3847/1538-4357/ab2da3}

\bibitem[{{Rappazzo} {et~al.}(2008){Rappazzo}, {Velli}, {Einaudi}, \&
  {Dahlburg}}]{rappazzo2008}
{Rappazzo}, A.~F., {Velli}, M., {Einaudi}, G., \& {Dahlburg}, R.~B. 2008, \apj,
  677, 1348, \dodoi{10.1086/528786}

\bibitem[{{Ruderman}(2017)}]{ruderman2017}
{Ruderman}, M.~S. 2017, \solphys, 292, 111, \dodoi{10.1007/s11207-017-1133-0}

\bibitem[{{Ruderman} \& {Terradas}(2015)}]{ruderman2015}
{Ruderman}, M.~S., \& {Terradas}, J. 2015, \aap, 580, A57,
  \dodoi{10.1051/0004-6361/201526168}

\bibitem[{{Shestov} {et~al.}(2017){Shestov}, {Nakariakov}, {Ulyanov}, {Reva},
  \& {Kuzin}}]{shestov2017}
{Shestov}, S.~V., {Nakariakov}, V.~M., {Ulyanov}, A.~S., {Reva}, A.~A., \&
  {Kuzin}, S.~V. 2017, \apj, 840, 64, \dodoi{10.3847/1538-4357/aa6c65}

\bibitem[{{Shoda} {et~al.}(2019){Shoda}, {Suzuki}, {Asgari-Targhi}, \&
  {Yokoyama}}]{shoda2019}
{Shoda}, M., {Suzuki}, T.~K., {Asgari-Targhi}, M., \& {Yokoyama}, T. 2019,
  \apjl, 880, L2, \dodoi{10.3847/2041-8213/ab2b45}

\bibitem[{{Shoda} \& {Yokoyama}(2018)}]{shoda2018b}
{Shoda}, M., \& {Yokoyama}, T. 2018, \apjl, 859, L17,
  \dodoi{10.3847/2041-8213/aac50c}

\bibitem[{{Spruit}(1981)}]{spruit1981}
{Spruit}, H.~C. 1981, \aap, 98, 155

\bibitem[{{Suzuki} \& {Inutsuka}(2005)}]{suzuki2005}
{Suzuki}, T.~K., \& {Inutsuka}, S.-i. 2005, \apjl, 632, L49,
  \dodoi{10.1086/497536}

\bibitem[{{Terradas} {et~al.}(2008){Terradas}, {Andries}, {Goossens},
  {Arregui}, {Oliver}, \& {Ballester}}]{terradas2008b}
{Terradas}, J., {Andries}, J., {Goossens}, M., {et~al.} 2008, \apjl, 687, L115,
  \dodoi{10.1086/593203}

\bibitem[{{Terradas} \& {Goossens}(2012)}]{terradas2012}
{Terradas}, J., \& {Goossens}, M. 2012, \aap, 548, A112,
  \dodoi{10.1051/0004-6361/201219934}

\bibitem[{{Terradas} \& {Ofman}(2004)}]{terradas2004}
{Terradas}, J., \& {Ofman}, L. 2004, \apj, 610, 523, \dodoi{10.1086/421514}

\bibitem[{Tomczyk {et~al.}(2007)Tomczyk, McIntosh, Keil, Judge, Schad, Seeley,
  \& Edmondson}]{tomczyk2007}
Tomczyk, S., McIntosh, S.~W., Keil, S.~L., {et~al.} 2007, Science, 317, 1192,
  \dodoi{10.1126/science.1143304}

\bibitem[{{van Ballegooijen} {et~al.}(2011){van Ballegooijen}, {Asgari-Targhi},
  {Cranmer}, \& {DeLuca}}]{vanballegooijen2011}
{van Ballegooijen}, A.~A., {Asgari-Targhi}, M., {Cranmer}, S.~R., \& {DeLuca},
  E.~E. 2011, \apj, 736, 3, \dodoi{10.1088/0004-637X/736/1/3}

\bibitem[{{van der Holst} {et~al.}(2014){van der Holst}, {Sokolov}, {Meng},
  {Jin}, {Manchester}, {T{\'o}th}, \& {Gombosi}}]{vanderholst2014}
{van der Holst}, B., {Sokolov}, I.~V., {Meng}, X., {et~al.} 2014, \apj, 782,
  81, \dodoi{10.1088/0004-637X/782/2/81}

\bibitem[{{Van Doorsselaere} {et~al.}(2018){Van Doorsselaere}, {Antolin}, \&
  {Karampelas}}]{vd2018}
{Van Doorsselaere}, T., {Antolin}, P., \& {Karampelas}, K. 2018, \aap, 620,
  A65, \dodoi{10.1051/0004-6361/201834086}

\bibitem[{{Van Doorsselaere} {et~al.}(2014){Van Doorsselaere}, {Gijsen},
  {Andries}, \& {Verth}}]{vd2014}
{Van Doorsselaere}, T., {Gijsen}, S.~E., {Andries}, J., \& {Verth}, G. 2014,
  \apj, 795, 18, \dodoi{10.1088/0004-637X/795/1/18}

\bibitem[{{Van Doorsselaere} {et~al.}(2008){Van Doorsselaere}, {Nakariakov}, \&
  {Verwichte}}]{vd2008}
{Van Doorsselaere}, T., {Nakariakov}, V.~M., \& {Verwichte}, E. 2008, \apjl,
  676, L73

\bibitem[{{Verbeke} {et~al.}(2019){Verbeke}, {Mays}, {Temmer}, {Bingham},
  {Steenburgh}, {Dumbovi{\'c}}, {N{\'u}{\~n}ez}, {Jian}, {Hess}, {Wiegand},
  {Taktakishvili}, \& {Andries}}]{verbeke2019}
{Verbeke}, C., {Mays}, M.~L., {Temmer}, M., {et~al.} 2019, Space Weather, 17,
  6, \dodoi{10.1029/2018SW002046}

\bibitem[{{Verdini} {et~al.}(2019){Verdini}, {Grappin}, \&
  {Montagud-Camps}}]{verdini2019}
{Verdini}, A., {Grappin}, R., \& {Montagud-Camps}, V. 2019, \solphys, 294, 65,
  \dodoi{10.1007/s11207-019-1458-y}

\bibitem[{{Wang} {et~al.}(2012){Wang}, {Ofman}, {Davila}, \& {Su}}]{wang2012}
{Wang}, T., {Ofman}, L., {Davila}, J.~M., \& {Su}, Y. 2012, \apjl, 751, L27,
  \dodoi{10.1088/2041-8205/751/2/L27}

\bibitem[{{Wentzel}(1979)}]{wentzel1979}
{Wentzel}, D.~G. 1979, \apj, 227, 319, \dodoi{10.1086/156732}

\bibitem[{{Yuan} \& {Van Doorsselaere}(2016)}]{yuan2016a}
{Yuan}, D., \& {Van Doorsselaere}, T. 2016, \apjs, 223, 23,
  \dodoi{10.3847/0067-0049/223/2/23}

\bibitem[{{Zaitsev} \& {Stepanov}(1975)}]{zaitsev1975}
{Zaitsev}, V.~V., \& {Stepanov}, A.~V. 1975, Issled. Geomagn. Aeron. Fiz.
  Solntsa, 3

\end{thebibliography}
\bibliographystyle{aasjournal}



\end{document}